\documentstyle[twoside,fleqn,espcrc2,epsf]{article}

\newcommand{\AmS}{{\protect\the\textfont2
  A\kern-.1667em\lower.5ex\hbox{M}\kern-.125emS}}

\newcommand{\bee}{\begin{equation}}
\newcommand{\ee}{\end{equation}}
\newcommand{\beea}{\begin{eqnarray}}
\newcommand{\eea}{\end{eqnarray}}

\title{Ginsparg-Wilson Games}

\author{Thomas DeGrand \\[2mm]
        Department of Physics, 
        University of Colorado\\ Boulder, CO 80309-390, USA}

\begin{document}

\begin{abstract}
I implement a set of tricks for constructing lattice fermion actions which
approximately realize the Ginsparg-Wilson relation, with very promising
results from simulations.
\end{abstract}

\maketitle

It might be useful to have a simple lattice fermion action
 $S= \bar \psi D \psi$ which approximately obeys the
Ginsparg-Wilson (GW) relation \cite{GW}
\bee
 \{ \gamma_5, D \} = D \gamma_5 R D.
\ee
Published algorithms\cite{EXACTGW}
 cost (apparently) hundreds of times as much as the
usual clover action. I describe an approach which costs about a 
factor of $6.5 \times (N+1)$ as much as the clover action 
for an $N$th order approximation, and even $N=1$ looks quite promising.

The ideas in this work are based on three remarkable formulas first
published by Neuberger\cite{NEUBERGER}:
Introducing a zeroth-order Dirac operator $D_0$
and defining $z = 1 - D_0/r_0$, a GW action (with $R=r_0$) is
\bee
 D_{GW} = r_0(1 - \frac{z}{\sqrt{z^\dagger z}}).
\ee
The inverse square root is approximated by 
\bee
 \frac{1}{\sqrt{z^\dagger z}} \simeq \frac{1}{N}
\sum_{j=1}^N \frac{1}{c_j z^\dagger z + s_j} 
\label{eq:ND}
\ee
($c_j=\cos^2(\pi(j+1/2)/(2N))$, $s_j=1-c_j$) and
\bee
 1 - \frac{A}{B} = \frac{B-A}{B}. 
\ee
Here $B-A = W^{(N)}$ is the polynomial
\bee
 \prod_{j=1}^N(c_j z^\dagger z + s_j) -
 (z/N)\sum_j \prod_{i\ne j}(c_i z^\dagger z +s_i).
\ee
To use this for propagators, note
$ D_{GW}^{(N)} \psi = \phi $ is
\bee
\psi = (D_{GW}^{(N)})^{-1} \phi = B (W^{(N)})^{-1} \phi.
\ee
(i.e. $\psi$ is found by inverting the simple differential operator $W$,
and then multiplying by the local operator $B$.)
Of course, one needs a $D_0$ for which 
 Eqn. \ref{eq:ND} works well for small $N$.

A good  $D_0$  should already be very chiral. This immediately
suggests that we begin with a fat link action--these actions are
already quite chiral as shown by their small mass renormalization and
 $Z_A \simeq 1$) \cite{FAT}.

\begin{figure}[h!tb]
\begin{center}
\leavevmode
\epsfxsize=70mm
\epsfbox{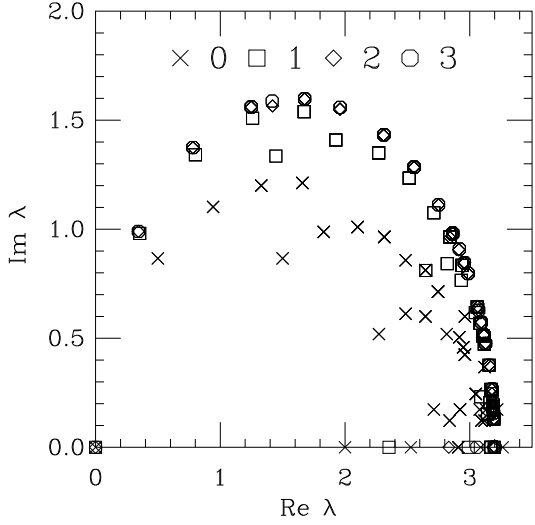}
\end{center}
\vspace{-28pt}
\caption{Free field  spectrum of $D_{GW}^{(N)}$,
with the planar action as the kernel, for $N=0$, 1, 2, 3.
Only the  Im $\lambda>0$ eigenvalues are shown. }
\label{fig:t3}
\end{figure}

The eigenvalues of a GW action  lie on a circle. I determine
the best $D_0$ by taking a free field test action and
 varying its parameterization to optimize its eigenvalue spectrum
 (in the least-squares
sense) for circularity, for some $r_0$
(which can also be varied; the optimal value is
about 1.6).  The action of choice is ''planar:'' it has
scalar and vector couplings
 $S=\sum_{x,r}\bar\psi(x)(\lambda(r) +i \gamma_\mu \rho_\mu(r))\psi(x+r)$
for $r$ connecting nearest neighbors ($\vec r=\pm\hat\mu$;
$\lambda=\lambda_1=-0.170$, $\rho_\mu=-0.177$) and diagonal
neighbors ($\vec r=\pm\hat\mu \pm\hat\nu$, $\nu\ne\mu$;
$\lambda=\lambda_2= -0.061$, $\rho_\mu= -0.053$;
 $\lambda(r=0)= -8\lambda_1 -24 \lambda_2$). The approach of the
eigenvalues to a circle is shown in Fig. \ref{fig:t3}.
The massive action for bare mass $m$  is obtained from the $m=0$
one by\cite{FERENC}
$D(m) = (1 + am/2) D_0 + am$.

\begin{figure}[h!tb]
\begin{center}
\leavevmode
\epsfxsize=70mm
\epsfbox{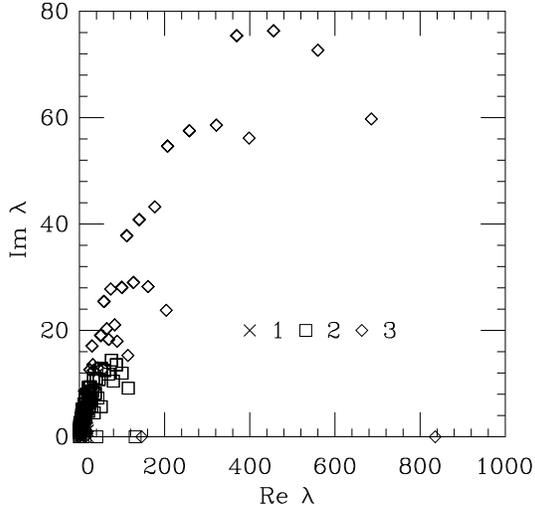}
\end{center}
\vspace{-28pt}
\caption{Free field  spectrum of the operator $W^{(N)}$,
with the Wilson action as the kernel, for $N=1$, 2, 3. }
\label{fig:wb}
\end{figure}

One might think that the iteration could be done starting with the
Wilson or clover
action. The  trick of Eqn. \ref{eq:ND}
 does rapidly pull the eigenmodes onto a circle,
but the problem is the decomposition into the $W/B$ form. Unlike for the
planar action, the
 eigenvalues of $W$ are thrown far out into the complex plane.
This is shown in Fig. \ref{fig:wb}.
Unfortunately, $W$ is the matrix which is to be inverted for
propagators.
Since the high momentum modes of a fat link action don't see the
gauge fields very well, they behave like free field modes. The wide
spread of eigenvalues means that in
real simulations, $W^{(N)}$ becomes ill-conditioned even
for small $N$.

\begin{figure}[h!tb]
\begin{center}
\leavevmode
\epsfxsize=70mm
\epsfbox{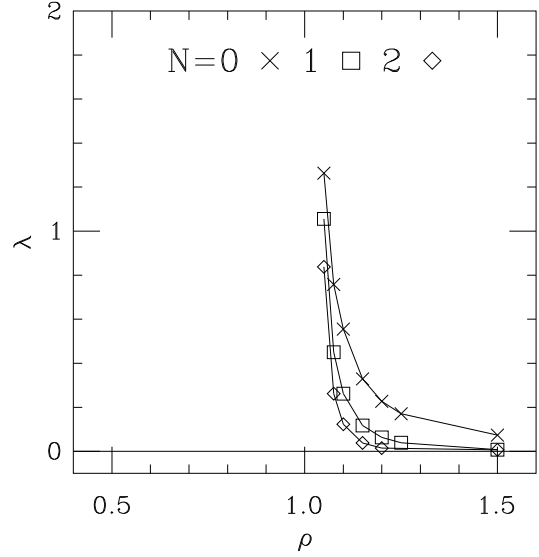}
\end{center}
\vspace{-28pt}
\caption{Real eigenvalue spectrum of $D_{GW}^{(N)}$ on background instanton
configurations, for $N=0$, 1, 2. }
\label{fig:mvrho}
\end{figure}

Chiral properties of the action,  in four dimensions,
are tested first by computing the value of its smallest real
 eigenvalue $\lambda$
 on a set of isolated instanton configurations
(the instanton radius is $\rho$) (Fig. \ref{fig:mvrho}).
In an exact GW action the real eigenvalue would be zero until
the instanton fell through the lattice, when it would
disappear. In an ordinary action, $\lambda$ is a 
smooth function of $\rho$, close to zero for big $\rho$ and moving away from 
zero, generally to a positive value, as $\rho$ decreases,
 until the eigenvalue collides with a doubler and goes
imaginary. For a better action, $\lambda$ keeps closer to zero and
breaks away more steeply, with
 a step function for $\lambda$ as the desired limiting result.

\begin{figure}[h!tb]
\begin{center}
\leavevmode
\epsfxsize=70mm
\epsfbox{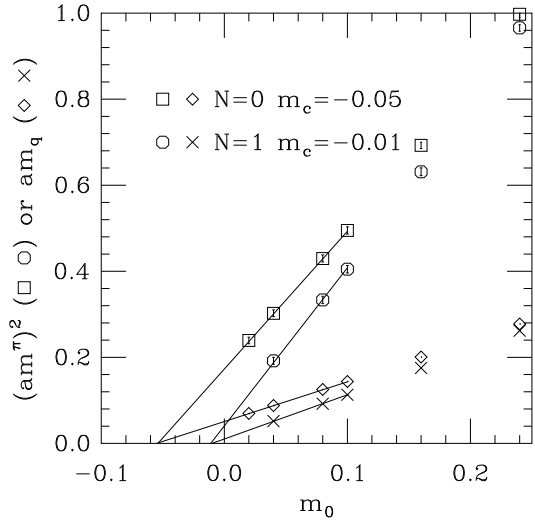}
\end{center}
\vspace{-28pt}
\caption{Pion mass and quark mass (from the PCAC relation) for $N=0$, 1. }
\label{fig:pisqb}
\end{figure}

As shown by the pion
 mass in Fig. \ref{fig:pisqb} (quenched,  for $SU(3)$,
$a=0.2$
 fm, $8^3\times 24$ lattice),the zeroth order action is already very chiral
and $N=1$ iteration is even more so.

\begin{figure}[h!tb]
\begin{center}
\leavevmode
\epsfxsize=70mm
\epsfbox{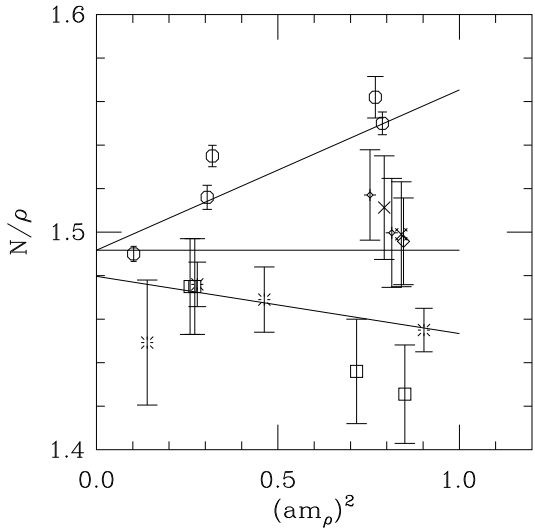}
\end{center}
\vspace{-28pt}
\caption{$N/\rho$ mass ratio at $\pi/\rho=0.7$.
  $N=0$ and $N=1$ actions  at $a= 0.2$
 fm are fancy crosses,
octagons--KS fermions,
bursts are NP $C_{SW}$ thin link actions, squares fat link clover on
Wilson gauge configurations, and the rest on improved gauge
 configurations. }
\label{fig:ratio}
\end{figure}

In Fig. \ref{fig:ratio} I show the $N/\rho$ mass ratio
at $\pi/\rho=0.7$ for the $N=0$ and 1 versions of this action,
along with other actions. Both of the new actions (on improved
background gauge configurations) have small scaling violations
 for this observable.

As an added feature, the hadron dispersion relation for these actions is
 better than for the
clover action (the extra terms in the planar action can be tuned to optimize
this\cite{PEREZ}).

Finally, a rough calculation of $Z_A$ from Ward identities produces a value
quite close to unity--as the fat link clover action\cite{FAT} gave.

To conclude: this is an approach towards the construction of a GW action
in which all the cost is ``up-front'' in the evaluation of $D_0$, but
the gain is that probably only a few terms 
(maybe just $N=0$ or 1) in the expansion of $D_{GW}^{(N)}$
are needed.
One also only needs to invert the simple (but messy) differential
operator $W^{(N)}$ (no inverse inside an inverse is needed).
It would be very interesting to tackle the hard lattice problem of
$\langle \bar \psi \psi \rangle$ along the lines of Ref. \cite{LELLOUCH}
with this approach.

This work was supported by the US Department of Energy.


\begin{thebibliography}{9}

\bibitem{GW}
P.~Ginsparg and K.~Wilson, Phys. Rev. D25 (1982) 2649.

\bibitem{EXACTGW}
Cf.
R.~Edwards, U.~M.~Heller, and R.~Narayanan, hep-lat/9807017,
P.~Hernandez, K.~Jansen and M. L\"uscher, hep-lat/9808010,
and A.~Borici, hep-lat/9810064.



\bibitem{NEUBERGER}
 H.~Neuberger,
Phys. Rev. Lett. 81 (1998) 4060;
hep-lat/9901003.

\bibitem{FAT}
  T.~DeGrand, et al, Nucl. Phys. B547 (1999) 259. See also
C.~Bernard and T.~DeGrand, this volume.

\bibitem{FERENC}
F.~Niedermayer, Lattice '98, Nucl. Phys. B (Proc. Suppl.) 73 (1999) 105.


\bibitem{PEREZ}
F.~Perez, work in progress.

\bibitem{LELLOUCH}
P.~Hern\'andez, K.~Jansen and L.~Lellouch, hep-lat/9907022.


\end{thebibliography}
\end{document}